\documentclass[12pt, draftclsnofoot, onecolumn]{IEEEtran}
\usepackage{amsfonts}
\usepackage{amssymb}
\usepackage{stfloats}
\usepackage{graphicx}
\usepackage{subfigure}
\usepackage{amsmath}
\usepackage{array}
\usepackage{epstopdf}
\usepackage{graphicx} 
\usepackage{amsthm} 
\usepackage{algorithm}
\usepackage{setspace}
\usepackage{algpseudocode}
\usepackage{amsmath}
\usepackage{graphics}
\usepackage{epsfig}
\usepackage{setspace}
\usepackage{extarrows}
\usepackage[usenames]{color}

\begin{document}

\title{An Indoor Environment Sensing and Localization System via mmWave Phased Array}

\author{Yifei Sun, Jie Li, Tong Zhang, Rui Wang, \\
	Xiaohui Peng, Tony Xiao Han, Haisheng Tan
	\thanks{
		Y. F. Sun, J. Li, T. Zhang, R. Wang. Department of Electronic and Electrical Engineering, Southern University of Science and Technology, Shenzhen 518055, China (e-mail: sunyf2019@mail.sustech.edu.cn; lij2019@ma\-il.sustech.edu.cn; zhangt7@sustech.edu.cn; wang.r@sustech.edu.cn).\\\indent
		Y. F. Sun. Department of Computer Science, The University of Hong Kong, Hong Kong 999077,  China (e-mail: sunyf2019@mail.sustech.edu.cn).\\\indent
		X. H. Peng, T. X. Han. Huawei Technologies Co., Ltd., Shenzhen 518129, China (e-mail: pengxiaohui5@huawei.com; tony.hanxiao@huawei.com).\\\indent
		H. S. Tan. LINKE Lab, School of Computer Science and Technology, University of Science and Technology of China, Hefei 230026, China (e-mail: hstan@ustc.edu.cn).\\\indent
		Corresponding author: Rui Wang.
	}
}

\maketitle

{\bf\textit{Abstract---}
	An indoor layout sensing and localization system in 60GHz millimeter wave (mmWave) band, named mmReality, is elaborated in this paper. The mmReality system consists of one transmitter and one mobile receiver, each with a phased array and a single radio frequency (RF) chain. To reconstruct the room layout, the pilot signal is delivered from the transmitter to the receiver via different pairs of transmission and receiving beams, so that the signals at all antenna elements can be resolved. Then, the spatial smoothing and two-dimensional multiple signal classification (MUSIC) algorithm is applied to detect the angle-of-arrival (AoAs) and angle-of-departure (AoDs) of the rays from the transmitter to the receiver. Moreover, the technique of multi-carrier ranging is adopted to measure the distance of each propagation path. Synthesizing the above geometrical parameters, the location of receiver relative to the transmitter can be pinpointed, both line-of-sight (LoS) and non-line-of-sight (NLoS) paths can also be determined. Therefore, the room layout can be reconstructed by moving the receiver and repeating the above measurement in different locations of the room. At the end, we show that the reconstructed room layout can be utilized to locate a mobile device according to its AoA spectrum, even with single access point.}



\section{Introduction}
Millimeter wave (mmWave) communications have been one of the key technologies of next generation wireless networks. Despite the large bandwidth, the high propagation and reflection loss of mmWave signals are the drawbacks from the communication point of view. However, these drawbacks may favor the wireless sensing performance by degrading the interference. Hence, it is of significant interest to exploit the sensing capability of mmWave communication system, such that the link reliability can be improved. For example, with the location knowledge of reflectors and mobile devices, the mmWave link can be quickly recovered if link blockage occurs.

There have been a few testbeds in the existing literature designated to detect the room layout via wireless transceiver. For example in \cite{zhu2015reusing}, a pair of mmWave transmitter and receiver, were deployed at a mobile platform to detect the layout of a corridor along a planned trajectory. In \cite{barneto2021millimeterwave}, a 5G mm-wave indoor mapping system was proposed, which utilizes orthogonal frequency division multiplexing (OFDM) radar processing to obtain sparse range–angle charts. However, the above designs may not be implemented in a wireless communication system, as base station (or access point) is usually deployed at a fixed location. It was shown in \cite{201466} that indoor ambient reflectors can be detected via a pair of mmWave transmitter and receiver, where the transmitter is fixed  and the receiver receive signals at multiple locations respectively. In this testbed, an omni-directional antenna on a rotation platform is used to simulate the receiving signal of multiple antennas, such that the AoAs and AoDs of propagation paths can be estimated. Moreover, based on detected layout, the SNR distribution in the room is also predicted.  However, it is more practical to use the phased array in mmWave communication systems, and imperfect antenna elements in phased array may degraded the estimation of AoA and AoD.

In order to quickly recover mmWave communication link from blockage, it is necessary for the BS to know both room layout and the location of the mobile device. Although a device can be localized in the procedure of layout reconstruction as in \cite{201466}, the signaling and time overhead are significant. Hence, a fast localization method exploiting the knowledge of layout is preferred. There have been a  number of works on the device localization via the AoA spectrum or received signal strength indicator (RSSI) fingerprint \cite{xiong2013arraytrack,10.1145/2829988.2787487,zhang2022toward}. For example, in \cite{xiong2013arraytrack}, exploiting the multiple-input-multiple-output (MIMO) technique and the prior location knowledge of the multiple access points (APs), a client can be localized by AoA spectra of APs. In \cite{10.1145/2829988.2787487}, the  direct-path AoA and RSSI were utilized for localization in a multi-AP system. However, all these localization methods require multiple APs. With single AP,  both angle and distance information of propagation paths were used for localization in \cite{10.1145/3210240.3210347,xie2019md}. In fact, due to the specular signal reflection in mmWave band, localization with one AP and room layout could be feasible. Thus, the cost of path distance measurement can be saved.

In this paper, an indoor sensing and localization system working in 60GHz band, namely mmReality, is proposed. The mmReality consists of one transmitter and one receiver, each with one phased array. The system works in two stages, namely layout reconstruction stage and localization stage. In the first stage, the AoAs, AoDs and lengths of the propagation paths between the transmitter and the receiver is estimated, where the receiver is put in multiple positions. Based on the above geometric parameters, the walls of the room can be detected and the layout can be reconstructed. In the second stage, the AoA spectrum for each indoor position can be predicted via the room layout, and the position of mobile device can be detected by comparing the its AoA spectrum with the prediction.

The remainder of this paper is organized as follows. In Section II, the architecture of the mmReality is introduced. In Section III, the estimation methods for AoAs, AoDs and lengths of propagation paths are elaborated. In Section IV, the algorithm for layout reconstruction is introduced. In Section V, the localization method based on room layout and AoA spectrum is discussed. The experiment results are elaborated in Section VI and the conclusion is drawn in Section VII.

\section{System Overview}
\begin{figure}[tb]
	\centering
	\includegraphics[width=0.45\linewidth]{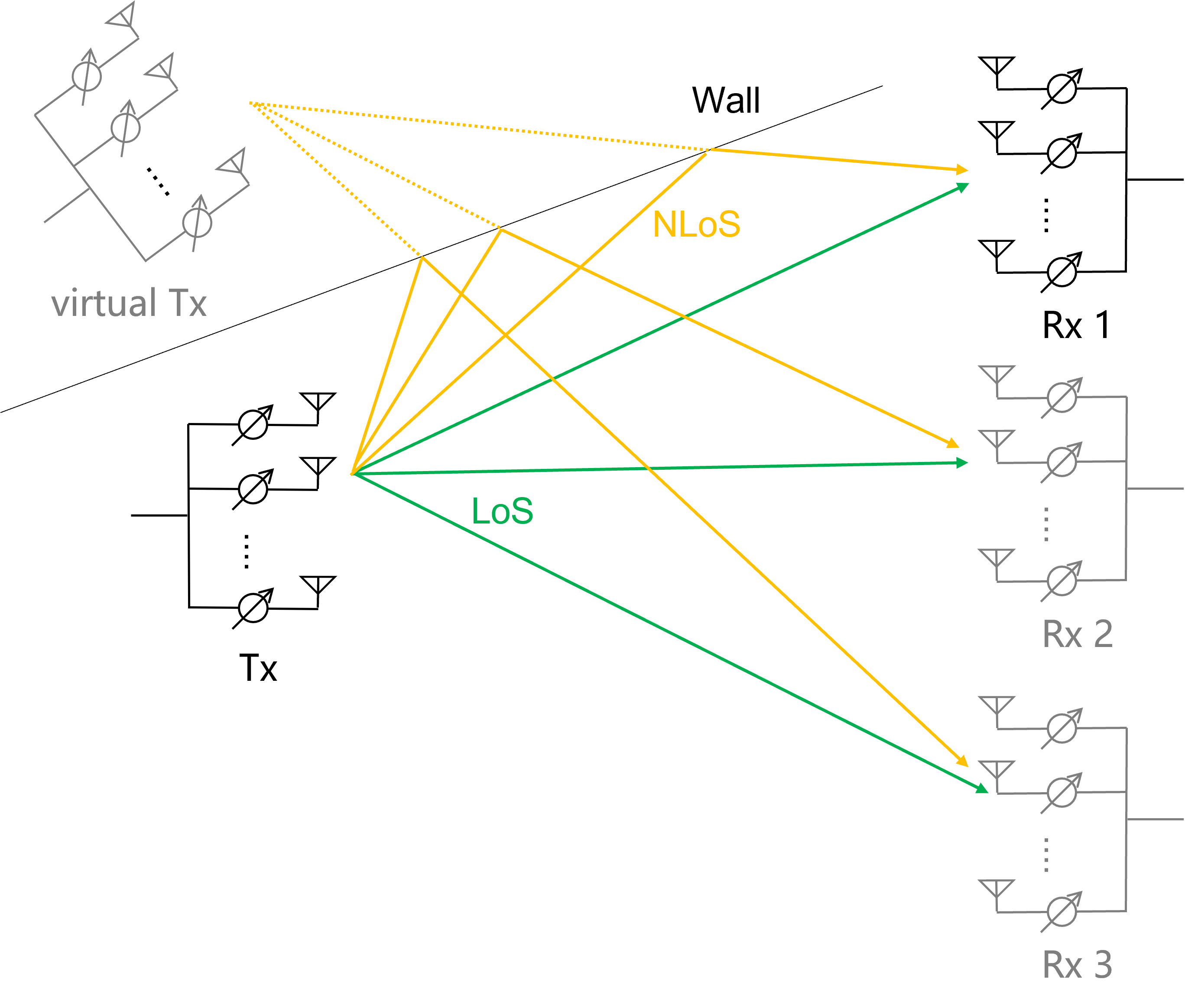}
	\caption{Illustration of propagation channel between the transmitter and the receiver, where the transmitter is fixed and a virtual transmitter can be observed by the receiver due to the specular reflection of the wall. }
	\label{fig:system_overview}
\end{figure}

The proposed mmReality system consists of one transmitter at a fixed location of a room and one mobile receiver, each with a phased array working at the 60GHz and one RF chain. There are $N_{\mathrm{T}}$ transmit antenna elements (TAEs) and $N_{\mathrm{R}}$ receive antenna elements (RAEs) in the phased arrays respectively. The propagation paths from  the transmitter to the receiver, including the line-of-sight (LoS) and non-line-of-sight (NLoS) paths, are illustrated in Fig. \ref{fig:system_overview}, where the specular reflection via walls dominates the NLoS paths. Via the multi-carrier pilot signal and the phased array, the transmitter and receiver can cooperatively estimate the AoAs, AoDs and lengths of propagation paths, including the LoS and NLoS paths. Hence, the relative locations of the main reflectors (walls) and the mobile receiver with respect to the transmitter can be detected via the geometric relations. The receiver is patrolling in the room, so that a complete picture of layout can be reconstructed. In practice, the transmitter can also communicate with multiple receivers at different locations of the room to complete the layout detection. In the following Section \ref{sec:3}, we first introduce the detection algorithms for AoA, AoD and path length, then the layout reconstruction algorithm will be elaborated in Section \ref{sec:reconstruction}.

\section{Estimation of AoA, AoD and Path Length}
\label{sec:3}
\subsection{Joint AoA and AoD Estimation}
The detection of AoA and AoD is challenging with analog MIMO architecture. For example, the exhaustive beam search used in \cite{barneto2021millimeterwave} is with high estimation overhead and low angular resolution. Although there have been a significant amount of research efforts spent on angle detection in multi-antenna systems, e.g., the Multiple Signal Classification (MUSIC) algorithm and Estimating Signal Parameter via Rotational Invariance Techniques (ESPRIT) algorithm, they are designed for digital MIMO systems, where the signals at all the antenna elements can be sampled individually. Notice that in the scenario of room layout sensing, the main reflectors (e.g., wall) are static, the propagation channel can be treated as quasi-static during multiple transmissions. This has been verified and utilized in \cite{201466} for the AoA and AoD detection with single rotating antenna. In this paper, exploiting this phenomenon, the transmitter can duplicate pilot transmissions via different transmission beams, while the receiver can adopt different receiving beams, so that the signals at RAEs can be resolved.

Specifically, the pilot signal $ s[t] $ ($ t=1,2,...,N_S $) is periodically transmitted via $ N_T $ different analog precoders, namely $ \{\mathbf{f}_1, \mathbf{f}_2,...,\mathbf{f}_{N_T}\} $, where $ N_S $ is the number of symbols in the transmission signal. For each analog precoder, the signal $ s[t] $ is transmitted for $ N_R $ times, and $ N_R $ different analog combiner is used at the receiver, namely $ \{\mathbf{w}_1, \mathbf{w}_2,...,\mathbf{w}_{N_R}\} $. For the elaboration convenience, we refer to the transmission with the $ i $-th precoder and $ j $-th combiner as the $ (i,j) $-th transmission. The received signal of the $(i,j)$-th transmission can be expressed as
\begin{eqnarray}
	\tilde{y}_{i,j}[t]&= &\mathbf{w}_{j}^{\mathsf{H}}\mathbf{H}\mathbf{f}_{i}s[t]+n[t]\\
	&=&\mathbf{w}_{j}^{\mathsf{H}}\mathbf{S}[t]\mathbf{f}_{i}+n[t],
\end{eqnarray}
where $\mathbf{w}_{j}^{\mathsf{H}}$ denotes the conjugate transpose of the $j$-th combiner $\mathbf{w}_{j}$, $\mathbf{H}$ is the channel matrix, $n[t]$ is the additive white Gaussian noise, and
\begin{equation}
	\mathbf{S}[t]=\mathbf{H}s[t]\in\mathbb{C}^{N_{\mathrm{R}}\times N_{\mathrm{T}}}
\end{equation}
denotes the TAE-to-RAE signal matrix. Aggregating the signal transmission via all precoder and combiner pairs, we have
\begin{equation}
	\widetilde{\mathbf{Y}}[t]=\mathbf{W}^{\mathsf{H}}\mathbf{S}[t]\mathbf{F}+\mathbf{N}[t],
\end{equation}
where the $(i,j)$-th entry of matrix $\widetilde{\mathbf{Y}}[t]$ is the received signal when applying the $i$-th precoder and the $j$-th combiner $\tilde{y}_{i,j}[t]$, $ \mathbf{F} = [\mathbf{f}_1, \mathbf{f}_2,...,\mathbf{f}_{N_T}] $, $ \mathbf{W}= [\mathbf{w}_1, \mathbf{w}_2,...,\mathbf{w}_{N_R}] $, and $ \mathbf{N}[t] $ is the aggregation of noise in all the transmissions. Applying the unitary beamforming matrix proposed in \cite{5379063}, we have $\mathbf{F}^{\mathsf{H}}\mathbf{F}=\mathbf{I}$ and $\mathbf{W}^{\mathsf{H}}\mathbf{W}=\mathbf{I}$. As a result, the signal matrix $\mathbf{S}[t]$ can be estimated from received signals $ \widetilde{\mathbf{Y}}(t) $ as
\begin{equation}
	\widehat{\mathbf{S}}[t]=\mathbf{W}\widetilde{\mathbf{Y}}[t]\mathbf{F}^{\mathsf{H}},
\end{equation}
where $\widehat{\mathbf{S}}[t]$ denotes the estimation of $\mathbf{S}[t]$.

Two-dimensional-MUSIC (2D-MUSIC) algorithm with spatial smoothing is used for AoA and AoD estimation from $\widehat{\mathbf{S}}[t] $. Although the spatial smoothing may lose some degree-of-freedom in the angle detection, it significantly suppresses the detection error due to source correlation \cite{17520}. In order to proceed the spatial smoothing with subarray sizes $N_{\mathrm{T}}^{\prime}$ and $N_{\mathrm{R}}^{\prime}$ at the transmitter and receiver respectively, we first define $\widehat{\mathbf{S}}_{i,j}[t]$ as the submatrix of $\mathbf{S}[t]$ by extracting the entries from the $i$-th row to $(i+N_{T}^{'}-1)$-th row and from the $j$-th column to $(j+N_{R}^{'}-1)$-th column, and $\widehat{\mathbf{s}}_{\mathrm{v}}^{i,j}[t]$ as the vectorization of $\widehat{\mathbf{S}}_{i,j}[t]$. Hence, the covariance matrix of $\widehat{\mathbf{s}}_{\mathrm{v}}^{i,j}[t]$ is given by
\begin{equation}
	\widehat{\mathbf{R}}_{i,j}=\frac{1}{T}\sum_{t=1}^{T}\widehat{\mathbf{s}}_{\mathrm{v}}^{i,j}[t]\Big[\widehat{\mathbf{s}}_{\mathrm{v}}^{i,j}[t]\Big]^{\mathsf{H}},
\end{equation}
where $\Big[\widehat{\mathbf{s}}_{\mathrm{v}}^{i,j}[t]\Big]^{\mathsf{H}}$ denotes the conjugate transpose of $\widehat{\mathbf{s}}_{\mathrm{v}}^{i,j}[t]$.
Then the spatial smoothed covariance matrix is defined as the average covariance matrix of each subarray,
\begin{align}
	\widehat{\mathbf{R}}_{\mathrm{ss}}=\frac{1}{M_{\mathrm{T}}M_{\mathrm{R}}}\sum_{i=1}^{M_{\mathrm{T}}}\sum_{j=1}^{M_{\mathrm{R}}}\widehat{\mathbf{R}}_{i,j},
\end{align}
where $M_{\mathrm{T}}=N_{\mathrm{T}}-N_{\mathrm{T}}^{\prime}+1$ and $M_{\mathrm{R}}=N_{\mathrm{R}}-N_{\mathrm{R}}^{\prime}+1$. With the estimated path number $K$, $\widehat{\mathbf{R}}_{\mathrm{ss}}$ is further decomposed via the eigenvalue decomposition as
\begin{align}
	\widehat{\mathbf{R}}_{\mathrm{ss}}=\mathbf{E}_{\mathrm{s}}\mathbf{D}_{\mathrm{s}}\mathbf{E}^{\mathsf{H}}_{\mathrm{s}}+\mathbf{E}_{\mathrm{n}}\mathbf{D}_{\mathrm{n}}\mathbf{E}^{\mathsf{H}}_{\mathrm{n}},
\end{align}
where $\mathbf{D}_{\mathrm{s}}$ and $\mathbf{D}_{\mathrm{n}}$ are diagonal matrices whose diagonal entries are the $K$ largest and the $(N_{\mathrm{T}}^{\prime}N_{\mathrm{R}}^{\prime}-K)$ smallest eigenvalues of $\widehat{\mathbf{R}}_{\mathrm{ss}}$, respectively; $\mathbf{E}_{\mathrm{s}}$ and $\mathbf{E}_{\mathrm{n}}$ are matrices composed of the eigenvectors of $\widehat{\mathbf{R}}_{\mathrm{ss}}$ correspondingly.

As a result, the 2D-MUSIC spatial spectrum function is given by
\begin{align}
	f(\phi,\theta)=\frac{1}{[\mathbf{a}_{\mathrm{R}}(\phi)\otimes\mathbf{a}_{\mathrm{T}}(\theta)]^{\mathsf{H}}\mathbf{E}_{\mathrm{n}}\mathbf{E}_{\mathrm{n}}^{\mathsf{H}}[\mathbf{a}_{\mathrm{R}}(\phi)\otimes\mathbf{a}_{\mathrm{T}}(\theta)]},
\end{align}
where $\otimes$ is the Kronecker product,
\begin{align}
	\mathbf{a}_{\mathrm{T}}(\theta)=[1,e^{-j2\pi\frac{d\sin\theta}{\lambda}},\ldots,e^{-j2\pi\frac{(N_{\mathrm{T}}^{\prime}-1)d\sin\theta}{\lambda}}]^{\mathsf{T}} \nonumber \\
	\mathbf{a}_{\mathrm{R}}(\phi)=[1,e^{-j2\pi\frac{d\sin\phi}{\lambda}},\ldots,e^{-j2\pi\frac{(N_{\mathrm{R}}^{\prime}-1)d\sin\phi}{\lambda}}]^{\mathsf{T}}
\end{align}
denote the transmission and receiving array responses, $d$ , $\lambda$ and $\mathbf{x}^{\mathsf{T}}$ denote the inter-element spacing, the wavelength and the transpose of vector $\mathbf{x}$ respectively. The $K$ highest peaks of $f(\phi,\theta)$ refers to the estimated AoDs and AoAs of the propagation paths.

\subsection{Path Length Estimation}

With the knowledge of AoAs and AoDs of propagation paths, signals from different paths can be separated from $\widehat{\mathbf{S}}[t]$ via the transmission and receiving beamforming, so that the length of each path can be estimated individually. Specifically, the  path length estimation of the $ i $-th path, whose AoA and AoD are denoted as $\phi_{i}$ and $\theta_{i}$ respectively, is elaborated below. Let $ \mathbf{A}_{\mathrm{T},i}$ be the aggregation of transmission array response vectors of all AoDs with the first column as $ \mathbf{a}(\phi_{i}) $, $ \mathbf{A}_{\mathrm{R},i}$ be the aggregation of receiving array response vectors of all AoAs with the first column as $ \mathbf{a}(\theta_{i}) $, $\mathbf{e}_{\mathrm{T},1}=[1,0,\ldots,0]^{\mathsf{T}}\in\mathbb\{0,1\}^{N_{\mathrm{T}}}$ and $\mathbf{e}_{\mathrm{R},1}=[1,0,\ldots,0]^{\mathsf{T}}\in\mathbb\{0,1\}^{N_{\mathrm{R}}}$, the transmission and receiving beamforming vector for the $ i $-th path are given by
\begin{equation}
	\mathbf{a}_{\mathrm{T},i}^{\mathrm{null}}=(\mathbf{A}_{\mathrm{T},i}\mathbf{A}_{\mathrm{T},i}^{\mathsf{H}})^{-1}\mathbf{A}_{\mathrm{T},i}\mathbf{e}_{\mathrm{T},1},
\end{equation}
and
\begin{equation}
	\mathbf{a}_{\mathrm{R},i}^{\mathrm{null}}=(\mathbf{A}_{\mathrm{R},i}\mathbf{A}_{\mathrm{R},i}^{\mathsf{H}})^{-1}\mathbf{A}_{\mathrm{R},i}\mathbf{e}_{\mathrm{R},1},
\end{equation}
respectively. The received signal of the $ i $-th path can be estimated as
\begin{align}
	s_{i}[t]=\mathbf{a}_{\mathrm{R},i}^{\mathrm{null}}\mathbf{S}(t)(\mathbf{a}_{\mathrm{T},i}^{\mathrm{null}})^{\mathsf{H}}.
\end{align}

OFDM ranging, also called multi-tone ranging\cite{194922}, is applied on $s_{i}[t]$ to estimate the length of the i-th path. In OFDM system, signals are modulated to subcarriers with different frequencies for transmission. For the same distance, the phases of the received signals of all subcarriers are different, which can be exploited to estimate the path lengths. We select equally-separated subcarriers for ranging. Without the consideration of sampling frequency offset (SFO) and packet detect delay, the estimated path length can be represented by
\begin{align}
	\label{eqn:d}
	d^{\star}=\mathop{\arg\max}_{d}\left|\sum_{i=1}^{L}\exp\Bigg(j\bigg(\varphi_{i}-\frac{2\pi f_{i}d}{c}\bigg)\Bigg)\right|.
\end{align}
where $\varphi_{i}$ is the phase of the $i$-th subcarrier, $\frac{2\pi f_{i}d}{c}$ denotes the range-dependent phase offset of the $i$-th subcarrier, $f_{i}$ denotes the frequency of the $i$-th subcarrier, $c$ denotes the speed of light,  and $L$ is the number of subcarriers. As a remark notice that, when $d$ equals the ground truth, the range-dependent phase offsets of all subcarriers will be equal, i.e., their complex exponentials are in-phase.

In practice, however, the SFO and packet detect delay are not negligible\cite{9442375}. As in \cite{201466}, a reference calibration scheme can be used to address this issue. Suppose the measured phase is $\varphi_{i,ref}$ for the $i$-th subcarrier at a known distance $d_{0}$, then equation \eqref{eqn:d} can be reformulated as
\begin{align}
	\label{eqn:d_ref}
	d^{\star}=\mathop{\arg\max}_{d}\left|\sum_{i=1}^{L}\exp\Bigg(j\bigg(\varphi_{i}-\varphi_{i,\mathrm{ref}}-\frac{2\pi f_{i}(d-d_{0})}{c}\bigg)\Bigg)\right|.
\end{align}

\section{Reconstructions of Room Layout}
\label{sec:reconstruction}

With the technique introduced in the previous section, the geometrical parameters of propagation paths, including AoAs, AoDs and lengths, can be estimated. In this section, following the method introduced in \cite{201466}, we first locate the mobile receiver in each measurement (which is referred to as the measurement points in the remaining of this paper), and then reconstruct the room layout.

\subsection{Localization of measurement points}
\label{subsec:localization}

Since the magnetometer has been widely adopted in mobile devices, it is assumed that both the transmitter and receiver are able to share the same direction as the reference direction of AoA and AoD measurements. Moreover, the estimated geometrical parameters of paths are reported to the transmitter for the layout reconstruction. For the elaboration convenience, we treat the position of the transmitter as the origin of the coordinate system, define $\mathbf{p}_{i}^{\mathbf{rx}}$ as the coordinates of the $i$-th measurement point, $\mathbf{p}_{i,\ell}^{\mathrm{ref}}$ as the coordinates of the reflection point of $\ell$-th path at the $i$-th measurement point. Moreover, tracing back the AoA of one NLoS path (say the $\ell$-th path) from the $i$-th measurement point, the receiver can find the mirror position of the transmitter (virtual transmitter) with respect to a wall as illustrated in Fig. \ref{fig:system_overview}, which is denoted as $\mathbf{p}_{i,\ell}^{\mathrm{vtx}}$. For the elaboration convenience, we refer to the above mirrors of transmitter as the virtual transmitters. Suppose that there are $L$ paths found at the $i$-th measurement point $\mathbf{p}_{i}^{\mathbf{rx}}$, and the estimated AoA, AoD and distance of the $\ell$-th path are $(\widehat{\theta}_{i,\ell},\widehat{\phi}_{i,\ell},\widehat{r}_{i,\ell}),\ \ell=1,2,\ldots,L$. The detection of $\mathbf{p}_{i}^{\mathrm{rx}}$, denoted by $\widehat{\mathbf{p}}_{i}^{\mathrm{rx}}$  is elaborated below.

\begin{figure}[tb]
	\centering
	\includegraphics[width=0.45\linewidth]{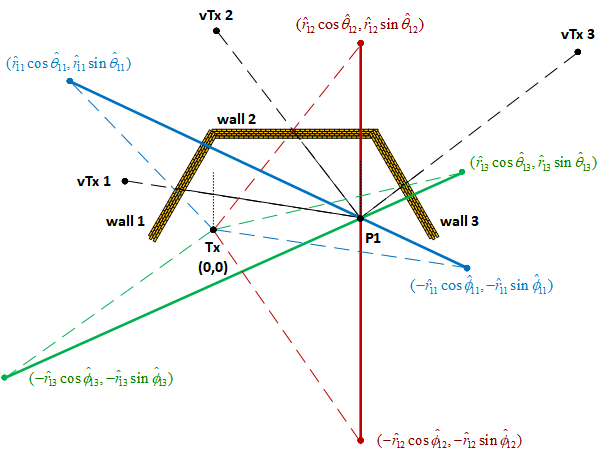}
	\caption{Localization of a measurement point, where vTx refers to the virtual transmitter.}
	\label{fig:localize_Rx}
\end{figure}

It can be proved that, without estimation error, measurement point must be on the line segment between $(\widehat{r}_{i,\ell}\cos\widehat{\theta}_{i,\ell},\widehat{r}_{i,\ell}\sin\widehat{\theta}_{i,\ell})$ and $(-\widehat{r}_{i,\ell}\cos\widehat{\phi}_{i,\ell},-\widehat{r}_{i,\ell}\sin\widehat{\phi}_{i,\ell})$, $(\ell=1,2,\ldots,L)$. An example is illustrated in Fig. \ref{fig:localize_Rx}, where the measurement point P1 is on the line segment connecting $(\widehat{r}_{1,1}\cos\widehat{\theta}_{1,1},\widehat{r}_{1,1}\sin\widehat{\theta}_{1,1})$ and $(-\widehat{r}_{1,1}\cos\widehat{\phi}_{1,1},-\widehat{r}_{1,1}\sin\widehat{\phi}_{1,1})$, which are marked by blue. Hence, as long as more than one path is detected in P1, its location can be estimated from the intersection points of corresponding line segments. In practice, due to measurement error, the intersection points are not unique, then the measurement point can be estimated by finding a point $\widehat{\mathbf{p}}_{i}^{\mathrm{rx}}=(\widehat{x}_{i}^{\mathrm{rx}},\widehat{y}_{i}^{\mathrm{rx}})$ that minimizes the sum distance to the line segments specified by all paths.

After pinpointing a measurement point, the estimated reflection points of the $\ell$-th path, denoted as $\widehat{\mathbf{p}}_{i,\ell}^{\mathrm{ref}}=(\widehat{x}_{i,\ell}^{\mathrm{ref}},\widehat{y}_{i,\ell}^{\mathrm{ref}})\ (\ell=1,2,\ldots,L)$, can be calculated via the following equation system
\begin{equation}
	\label{eqn:point_ref}
	\left\{
	\begin{aligned}
		 & \frac{\widehat{x}_{i,\ell}^{\mathrm{ref}}}{\widehat{y}_{i,\ell}^{\mathrm{ref}}}=\tan\widehat{\theta}_{i,\ell},                                                           \\
		 & \frac{\widehat{x}_{i,\ell}^{\mathrm{ref}}-\widehat{x}_{i}^{\mathrm{rx}}}{\widehat{y}_{i,\ell}^{\mathrm{ref}}-\widehat{y}_{i}^{\mathrm{rx}}}=\tan\widehat{\phi}_{i,\ell}.
	\end{aligned}
	\right.
\end{equation}
Moreover, the mirror position of the transmitter along  the  $\ell$-th path at the $i$-th measurement point can be estimated by
\begin{align}
	\widehat{\mathbf{p}}_{i,\ell}^{\mathrm{vtx}}=\widehat{\mathbf{p}}_{i}^{\mathrm{rx}}+\widehat{r}_{i,\ell}\left[\cos\widehat{\phi}_{i,\ell},\sin\widehat{\phi}_{i,\ell}\right]^{\mathsf{T}}.
\end{align}
As illustrated in the example of Fig. \ref{fig:localize_Rx}, there are three virtual transmitters from the P1 point of view. As a remark notice that the positions of virtual transmitters depends only on the position of transmitter and the layout of walls, and hence, receiver at different measurement points should see the same set of virtual transmitters.

\subsection{Layout Reconstruction}

It can be observed from Fig. \ref{fig:localize_Rx} that, after the localization of virtual transmitters, the wall can be detected by drawing the perpendicular bisector between the transmitter and each virtual transmitter. In order to obtain the complete room layout, we can move the receiver and take measurement in a number of points, so that the virtual transmitters with respect to all the walls can be captured.

In real measurement, there are still two issues on the above reconstruction method. First, due to the measurement error, the locations of one virtual transmitter detected at different measurement points may not be identical. Moreover, in addition to the first-order path (the path with one reflection from the transmitter to the receiver), some higher order paths (the path with more than one reflection) may also be detected. As a result, the detected locations of virtual transmitters may be dispersed. To classify the estimated positions of virtual transmitters corresponding to the same wall, we use Density-Based Spatial Clustering of Applications with Noise (DBSCAN) method and take the cluster centroid as the estimated position of the virtual transmitters.

\begin{figure}[htb]
	\centering
	\subfigure[Layout with reflection point $A$ and $B$.]{
		\includegraphics[width=0.45\linewidth]{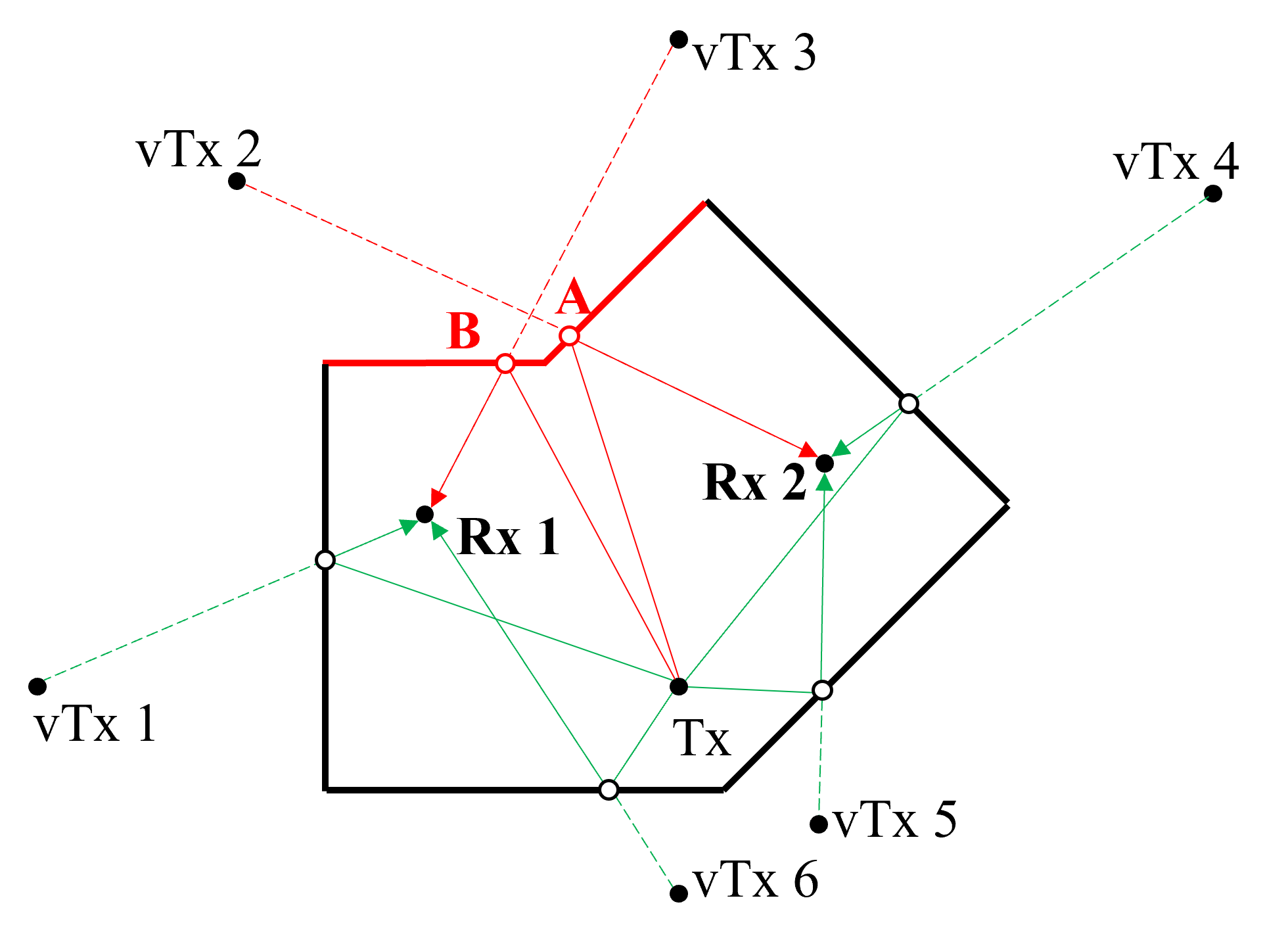}}\hfill
	\subfigure[Layout with reflection point $A'$ and $B'$.]{
		\includegraphics[width=0.45\linewidth]{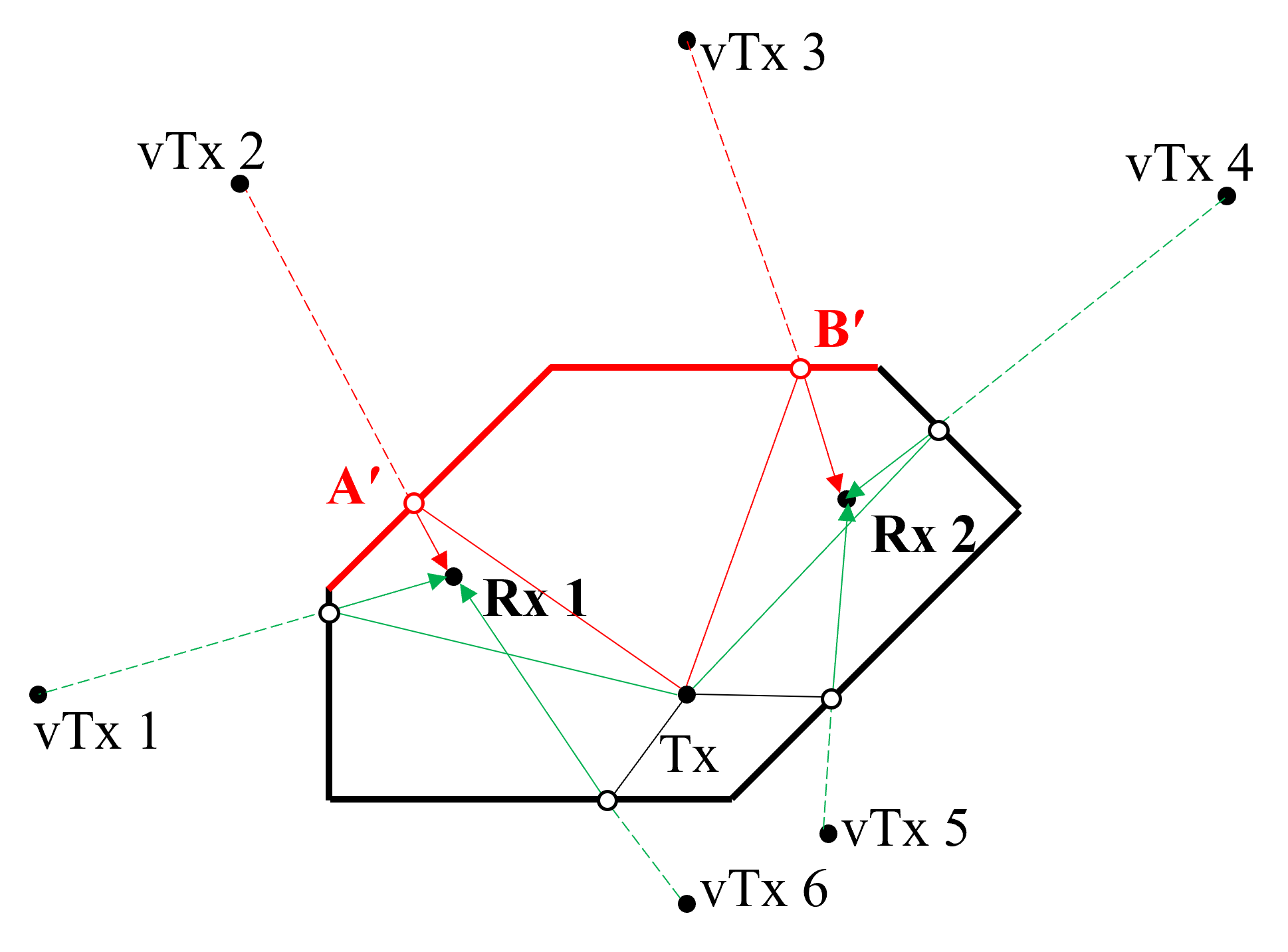}}
	\vspace{-0.2cm}
	\caption{Two possible layouts with the same virtual transmitter positions due to layout ambiguity, where vTx refers to the virtual transmitter.}
	\label{fig:layout_ambiguity}
\end{figure}

Finally, for some complicated room layouts, there are more than one possible room layouts given the locations of virtual transmitters. One example is illustrated in Fig. \ref{fig:layout_ambiguity}, where there are two possible layouts given the same locations of 6 virtual transmitters. This ambiguity can be removed by exploiting the locations of reflection points derived in \eqref{eqn:point_ref}. For example, in Fig. \ref{fig:layout_ambiguity}, if $A$ and $B$ are the reflection points of NLoS paths from vTx2 and vTx3 to the receivers respectively, then the room profile is shown in Fig. \ref{fig:layout_ambiguity}(a). Otherwise, if $A'$ and $B'$ are the reflection points, the room profile is shown in Fig. \ref{fig:layout_ambiguity}(b).

\section{Localization by AoA Spectrum}

In this section, the fast localization method for mobile receiver via AoA spectrum and reconstructed indoor layout is explained. Although the method introduced in Section \ref{subsec:localization} can be used to localize the receiver, the overhead is significant. First, the transmitter should deliver pilots dedicatedly to the receiver for at least $N_{\mathrm{T}}N_{\mathrm{R}}$ times, such that the AoA and AoD can be estimated with high resolution. However this estimation method may not be feasible when the channel is not quasi-static, e.g., there are moving persons in the room. Second, the offset between their local oscillators should be carefully calibrated, in order to suppress the estimation error of path length. The cost of oscillator synchronization could be high, especially in mmWave band.

In practice, the transmitter (i.e., BS or AP) would periodically broadcast control information to all the directions, so that each receiver can detect all the potential AoAs of the signals from the transmitter via periodic beam search. Hence, raw AoA estimation can be made by the receiver. By matching the observed dominant AoAs with the locations in the reconstructed room layout, the mobile receiver can be pinpointed. This facilitates the localization without dedicate signaling overhead. The existing works exploiting the AoA spectrum (including the AoAs and signal powers at the arrival directions) in localization usually rely on multiple APs. Compared with the existing works in sub-6GHz, we shall show single transmitter might be sufficient to generate the AoA spectruml for localization in mmWave band, with the assistance of virtual transmitters. The reasons are elaborated below. First, it is shown by experiments that the propagation paths arrived at the receiver are dominated by the LoS and first-order NLoS paths, leading to the limited number of virtual transmitters. Second, the phased array at the mmWave band is of smaller size, so that it can be implemented at mobile devices, and used to  resolve the real and virtual transmitters at different directions.

Specifically, we choose $g$ different positions in the reconstructed room layout, denoted as $\mathcal{G} =\{1,2,\ldots,g\}$, and calculate the AoAs at these positions according to the positions of real and virtual transmitters. Let $\Phi_{i}=\{\phi_{i,j}|j = 1,2,\ldots\}$ be the set of AoAs for the position $i$, and $\widehat{\Phi}$ be the set of AoAs observed at the real receiver. We define the error function of an expected AoA profiles with respect to the measured one as
\begin{align}
	f(\Phi_{i},\widehat{\Phi})=\sum_{j=1}^{|\Phi_{i}|}\min\Bigg(\Big\{|\phi_{i,j}-\widehat{\phi}_{k}|^{2}\Big|k\Big\},A_{\mathrm{th}}^{2}\Bigg),
\end{align}
where $A_{\mathrm{th}}$ is a constant threshold, and $|\Phi_{i}|$ denotes the cardinality of the set $\Phi_{i}$. Then the estimated position with the measured AoA profile $\widehat{\Phi}$ can be obtained by
\begin{align}
	i^{\star}=\mathop{\arg\min}_{i\in\mathcal{G}}f(\Phi_{i},\widehat{\Phi}).
\end{align}
As a remark notice that the threshold $A_{\mathrm{th}}$ in (18) is to avoid significant difference in the error function $f(\cdot)$ when the desired LoS and first-order reflection paths are blocked or higher-order reflection paths are captured in the AoA detection.

\section{Experiment and Discussion}
\subsection{mmReality System Implementation}

\begin{figure}[tb]
	\centering
	\includegraphics[width=0.48\linewidth]{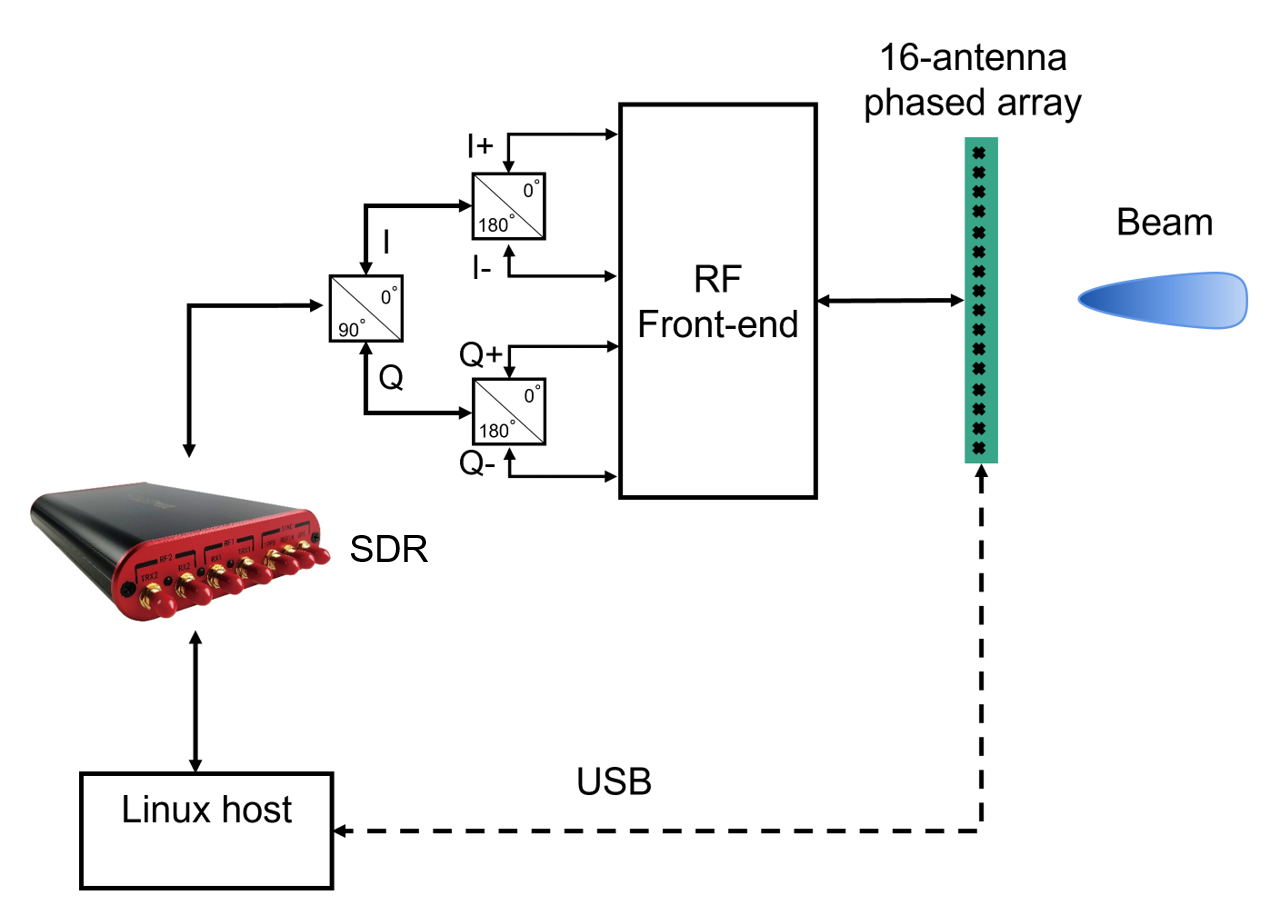}
	\caption{Hardware architecture.}
	\label{fig:architecture}
\end{figure}

A block diagram of mmReality is illustrated in Fig. \ref{fig:architecture}. At the transmitter side, one software defined radio (SDR) generates the baseband signal, and up-converts it to the intermediate frequency (IF) band centered at 500MHz. The IF signal is fed into the 90 degree and 180 degree power splitters sequentially to generate differential in-phase and quadrature (IQ) signals, and further up-converted to 60GHz by 16-antenna phased array. The antenna selection, beam codebook (direction) and power gain of the phased array is controlled by a host computer. Similarly, at the receiver side, the signal is captured and down-converted to the IF band by the phased array, and further down-converted to the baseband by the SDR. In order to facilitate the path length detection, the transmission signal is modulated with 8 subcarriers and $12.5$ MHz bandwidth via OFDM technology.

Before the measurements, the inter-element spacing and inter-element phase offset of the phased arrays should be calibrated, which is elaborated below.

\textbf{Inter-element spacing calibration:}
Although antenna arrays are designed with half-wavelength inter-element spacing to generate beam patterns with a single main lobe and low side lobes, the wavelength varies at different carrier frequencies. The difference of carrier frequencies in 60GHz band is generally more than 1GHz and cannot be neglected. Moreover, the direct measurement of inter-element spacing may not be accurate due to the short wavelength and small antenna size \cite{ayyalasomayajula2020locap}.

To address the above issue, we measure the inter-element spacing via the received signals of the phased array. We use a transmitter with horn antenna to transmit a single tone and mount the phased array to be calibrated on a rotation platform as the receiver. The mmWave absorbers are used to suppress the potential NLoS paths. The received signals are collected at $N_{\theta}$ different AoAs. At each AoA, the antenna elements are tuned on alternatively to receive the the same single tone. Let $\varphi_{m,n}$ be the the estimated phase of
the $n$-th antenna element at the $m$-th AoA, denoted as $\theta_{m}$, then the inter-element spacing can be estimated by
\begin{align}
	d^{\star}\!=\!\mathop{\arg\max}_{d}\sum_{n=2}^{N}\left|\sum_{m=1}^{N_{\theta}}\exp\Bigg(j\bigg(\varphi_{m,n}\!-\!\varphi_{m,n-1}\!-\!2\pi\frac{d\sin\theta_{m}}{\lambda}\bigg)\Bigg)\right|.
\end{align}

\textbf{Inter-element phase offset calibration:}
Notice that the phase offset is caused by the different transmission lines of antenna elements, we calibrate it with a transmitter at the boresight direction of the phased array. In the above calibration of inter-element spacing, let $\theta_1=0$, the phase offset difference of each two adjacent elements (say between the $n$-th and $n-1$-th antenna elements) due to transmission lines can be estimated directly by $\varphi_{1,n}-\varphi_{1,n-1}$. As a result, the phase offset of all the remaining antenna elements with respect to the first element can be compensated.

The mmReality system is deployed in a corridor with irregular layout, as illustrated in Fig. \ref{fig:corridor_layout}(a). The position of the transmitter is fixed, and the receiver is put at 15 measurement points respectively as in Fig. \ref{fig:corridor_layout}(a). At each measurement point, the OFDM-modulated signal is transmitted via 16 x 16 pairs of transmission and receiving beams, with a total duration of 8 s.

\begin{figure}[tb]
	\centering
	\includegraphics[width=0.5\linewidth]{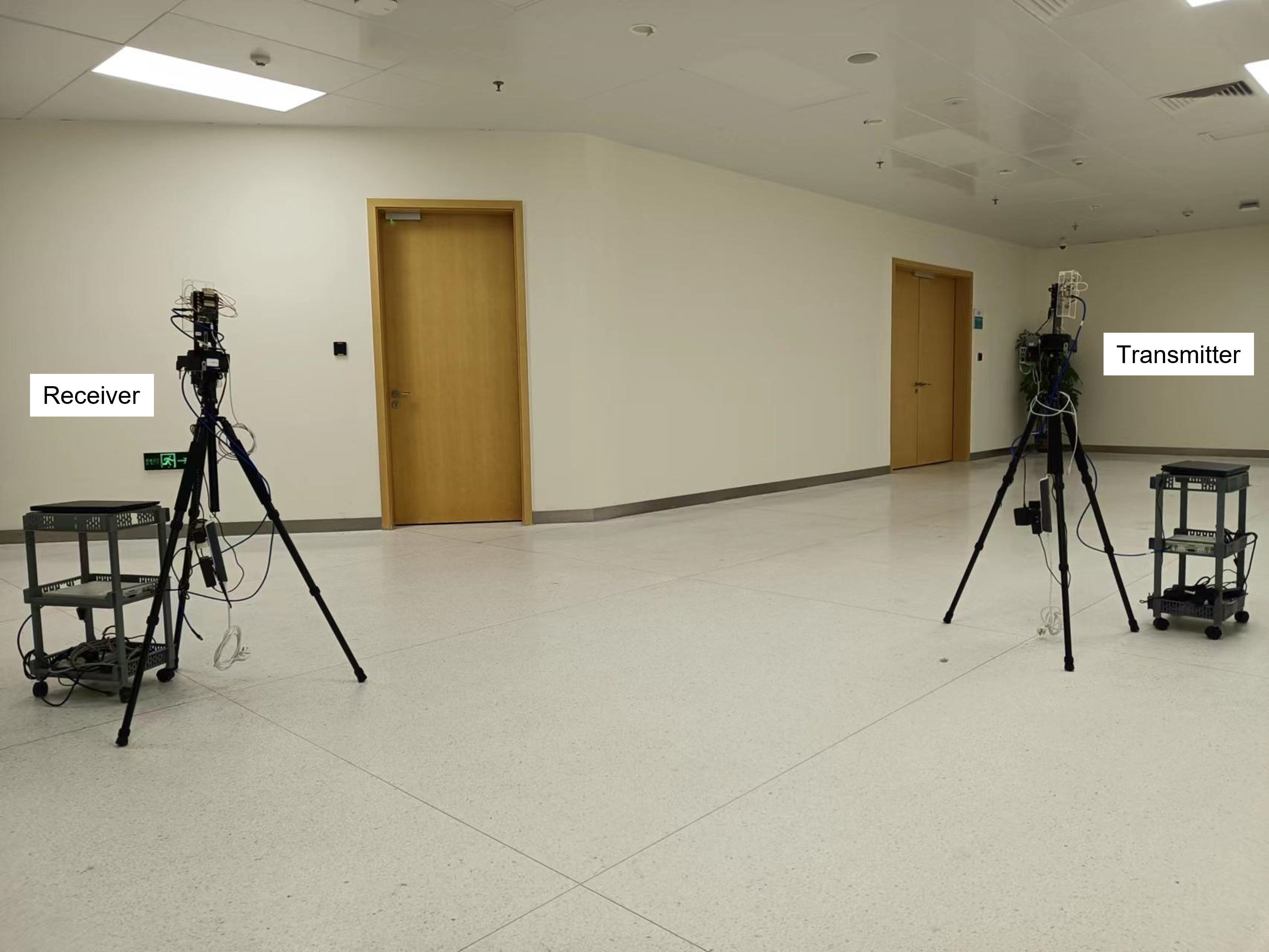}
	\caption{Experimental environment.}
	\label{fig:corridor }
\end{figure}

\begin{figure}[htb]
	\centering
	\subfigure[]{
		\includegraphics[width=0.45\linewidth]{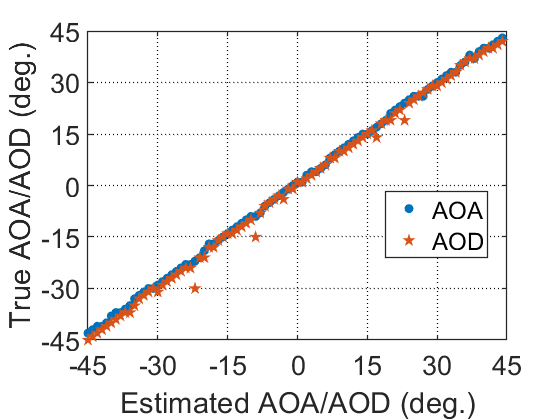}
	}\hfill
	\subfigure[]{
		\includegraphics[width=0.45\linewidth]{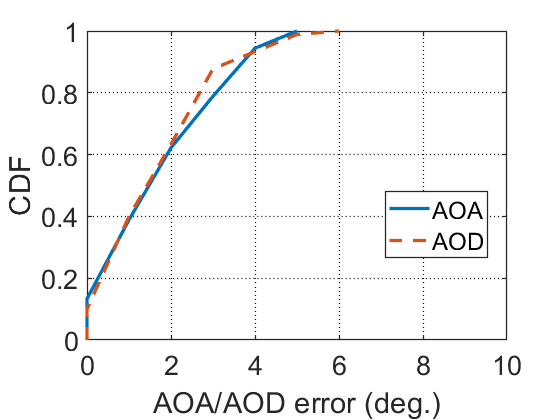}
	}\hfill
	\subfigure[]{
		\includegraphics[width=0.45\linewidth]{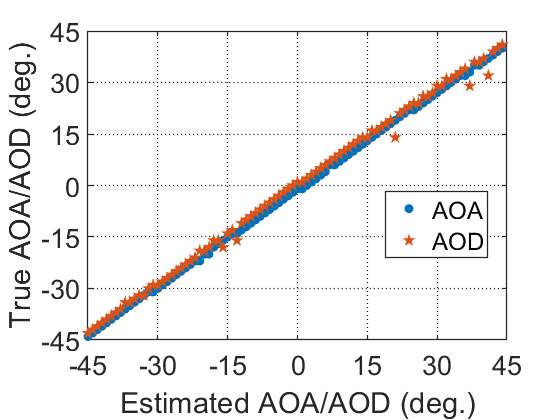}
	}\hfill
	\subfigure[]{
		\includegraphics[width=0.45\linewidth]{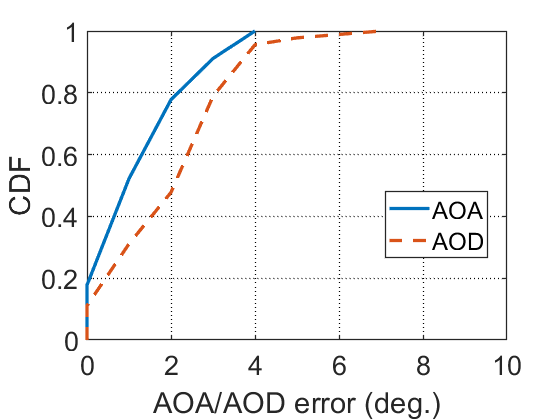}
	}
	\caption{Illustration of AoA/AoD estimation. (a) Estimated AoA/AoD versus ground truth in LoS scenario ; (b)  CDF of AoA/AoD estimation error in LoS scenario; (c) Estimated AoA/AoD versus ground truth in NLoS scenario; (d) CDF of AoA/AoD estimation error in NLoS scenario.}
	\label{fig:angle_error}
\end{figure}

\subsection{AoA/AoD Estimation}
In this part, the accuracy of AoA and AoD estimations is illustrated. We compare the estimated AoAs and AoDs with the ground truths, which are measured by a laser rangefinder in the real environment. In Fig. \ref{fig:angle_error}(a) and Fig. \ref{fig:angle_error}(c), the comparison between the estimated AoA and AoD and their true values are made, where the x-axis and y-axis represent the estimated values and the true ones  respectively. The estimated angles ranges from $-45^\circ$ to $45^\circ$, since the estimations at all the measurement points are included. The comparisons for LoS and NLoS paths are illustrated in Fig. \ref{fig:angle_error}(a) and Fig. \ref{fig:angle_error}(c), respectively. It can be observed that the esimated AoA/AoD matches true AoA/AoD in both LoS and NLoS scenarios. Moreover, the cumulative distribution functions (CDF) of the magnitudes of the estimation errors for both LoS and NLoS paths are illustrated in Fig. \ref{fig:angle_error}(b) and Fig. \ref{fig:angle_error}(d), respectively. It can be observed that 90\% of the estimation errors are less than  $4^\circ$.

\subsection{Path Length Estimation}
\begin{figure}[htb]
	\centering
	\subfigure[]{
		\includegraphics[width=0.45\linewidth]{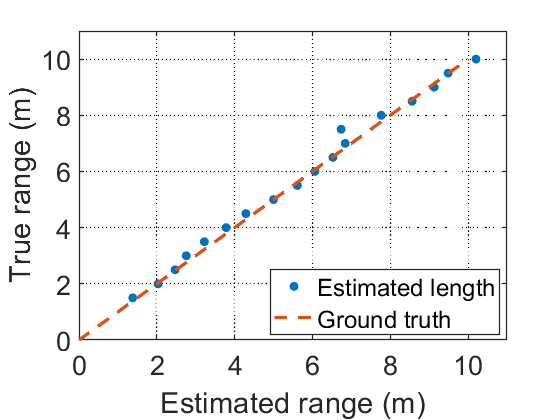}
	}\hfill
	\subfigure[]{
		\includegraphics[width=0.45\linewidth]{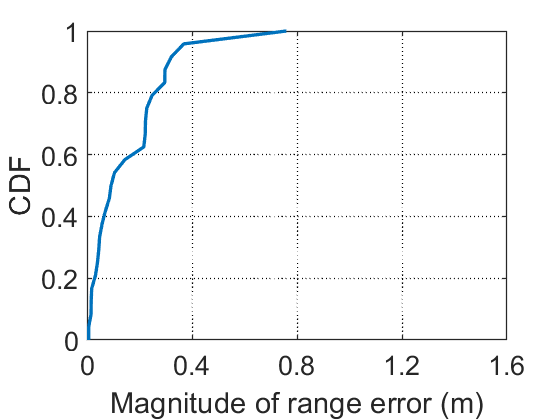}
	}\hfill
	\subfigure[]{
		\includegraphics[width=0.45\linewidth]{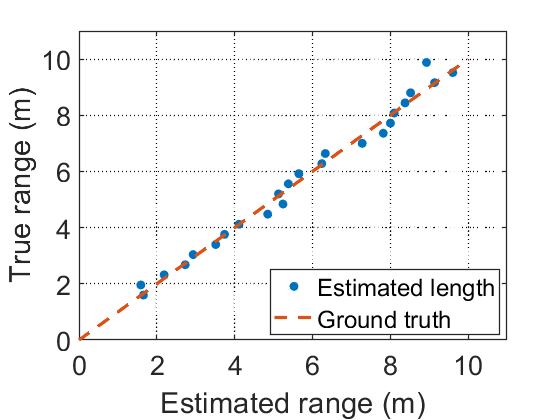}
	}\hfill
	\subfigure[]{
		\includegraphics[width=0.45\linewidth]{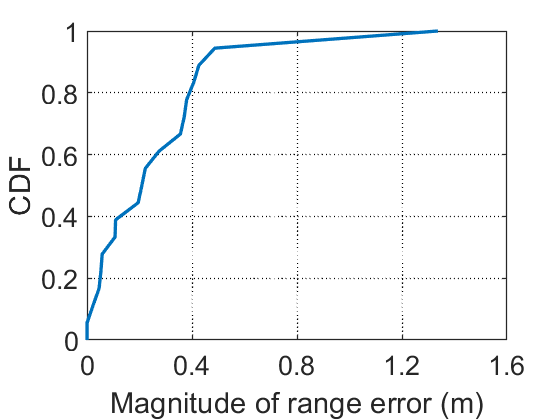}
	}
	\caption{Illustration of path length estimation. (a) Estimated path length versus ground truth in LoS scenario ; (b)  CDF of path length estimation error in LoS scenario; (c) Estimated path length versus ground truth in NLoS scenario; (d) CDF of path length estimation error in NLoS scenario.}
	\label{fig:range_error}
\end{figure}

In Fig. \ref{fig:range_error}(a) and Fig. \ref{fig:range_error}(c),  the comparison between the estimated path lengths and the true values are made, where the x-axis and y-axis represent the estimated values and the true ones. The estimated path lengths ranges from $1.6$ m to $10$ m, since the estimations at all the measurement points are included. It can be observed  that estimated path lengths matches true ones in both LoS and NLoS scenarios. The CDFs of the magnitudes of the estimation errors for both LoS and NLoS paths are illustrated in Fig. \ref{fig:range_error}(b) and Fig. \ref{fig:range_error}(d), respectively, where the average error magnitudes  for LoS and NLoS paths are $0.15$ m and $0.25$ m, respectively. The LoS paths  perform better than NLoS reflection paths due to higher SNR.

\subsection{Room Layout Reconstruction}
\begin{figure}[htb]
	\centering
	\subfigure[]{
		\includegraphics[width=0.48\linewidth]{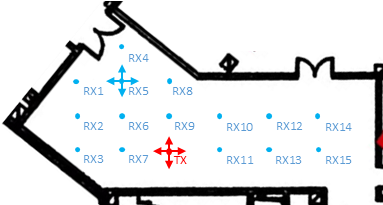}
	}\hfill
	\subfigure[]{
		\includegraphics[width=0.48\linewidth]{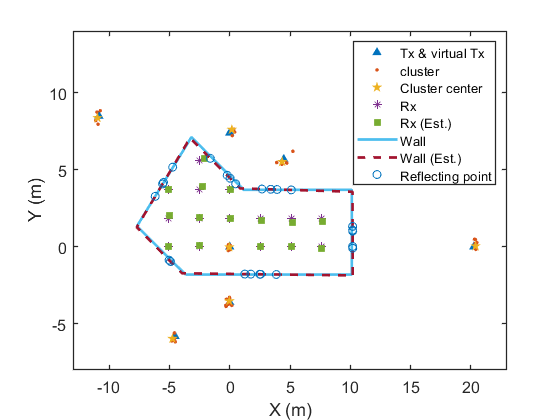}
	}
	\caption{(a) Corridor layout illustration with the positions of the transmitter and measurement points. (b) Corridor layout reconstruction.}
	\label{fig:corridor_layout}
\end{figure}

Integrating the estimated AoAs, AoDs and path lengths at all measurement points, the reconstructed layout of the corridor is illustrated in  Fig. \ref{fig:corridor_layout}(b), where the blue line and red line represent the real and estimated walls, respectively. It can be seen that all the 6 walls are detected with high estimation accuracy, as both lines almost overlaps. Moreover, the estimated positions of measurement points, virtual transmitters and reflecting points are also illustrated in Fig. \ref{fig:corridor_layout}(b). As a remark notice that higher-order reflections can be found in the measurements, although the LoS and first-order reflection paths are dominant. However, since the virtual transmitters of the higher-order reflection paths are sparse, they are eliminated by the DBSCAN algorithm.  Hence, only the estimated virtual transmitter positions of the first-order reflection is shown in Fig. \ref{fig:corridor_layout}(b).

\begin{figure}[htb]
	\centering
	\subfigure[]{
		\includegraphics[width=0.45\linewidth]{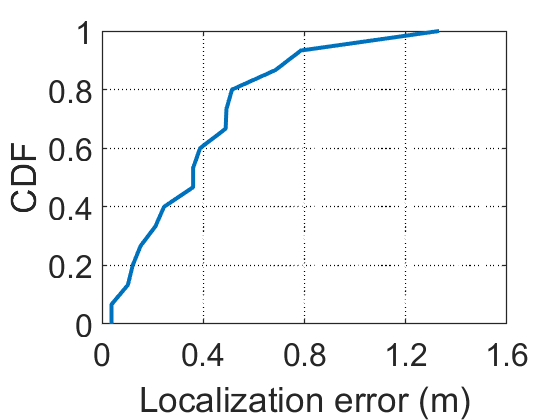}
	}\hfill
	\subfigure[]{
		\includegraphics[width=0.45\linewidth]{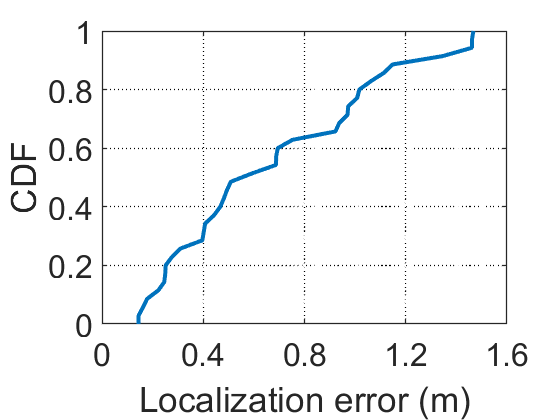}
	}
	\caption{Distribution of localization errors. (a) CDF of localization errors of measurement points. (b) CDF of localization errors of reflection points.}
	\label{fig:error_reconstruction}
\end{figure}

The CDFs of the localization errors of the measurement points and reflection points are illustrated in Fig. \ref{fig:error_reconstruction}(a) and Fig. \ref{fig:error_reconstruction}(b) , respectively. The average localization error of the measurement points is $0.42$ m, and 90\% of the localization errors are below $0.8$ m, while those of the estimated reflection points are $0.6$ m and $1.2$ m respectively. The localization error of reflection points is generally larger. This is because the localization of reflection points is based on the estimated positions of measurement points.

\subsection{Localization via AoA Spectrum}
Based on the detected corridor layout in the above experiments, the localization performance via AoA spectrum is demonstrated in this part. Specifically, the receiver is randomly put at 25 positions of the corridor respectively, and the AoAs at each position are measured by the receiver. According to the fast localization method introduced in Section V, the coordinates of each position can be detected, and the CDF of localization errors is illustrated in Fig. 12. It can be seen that 90\% of the localization error is within 1.3 m, and the average localization error is 1.0 m. One example of localization via AoA spectrum is shown in Fig. 11, where different colors are used to demonstrate the value of error function $f(\cdot)$ defined in (18). The estimated location of the receiver is determined by find a position in $\mathcal{G}$ with the smaller value of error function. The localization error is mainly due to the detection error of the positions of virtual transmitters.

\begin{figure}[htb]
	\centering
	\subfigure[]{
		\includegraphics[width=0.48\linewidth]{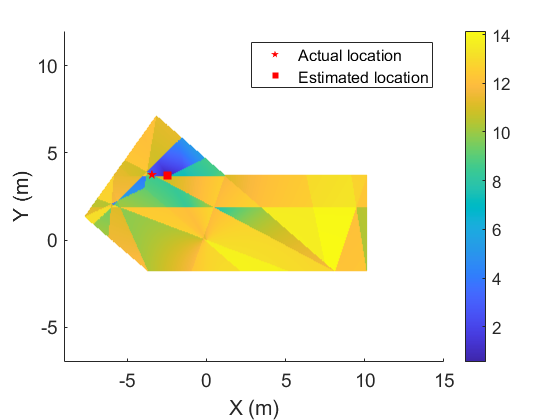}
	}\hfill
	\subfigure[]{
		\includegraphics[width=0.48\linewidth]{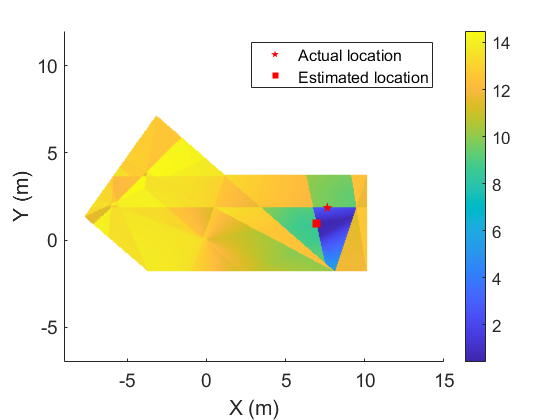}
	}
	\caption{Two examples of localization via AoA spectrum.}
	\label{fig:localization}
\end{figure}

\begin{figure}[h]
	\centering
	\includegraphics[width=0.48\linewidth]{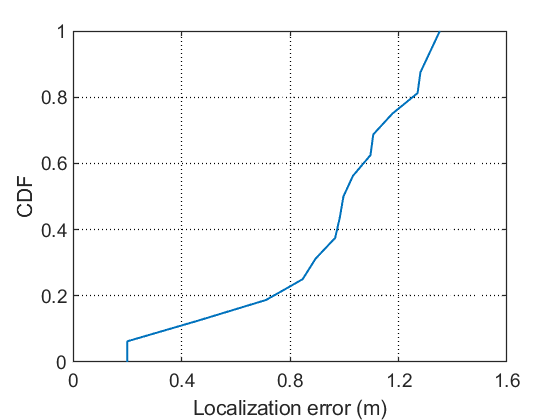}
	\caption{CDF of localization error.}
	\label{fig:error_CDF_localization}
\end{figure}

\section{Conclusion}
In this paper, the mmReality system for indoor layout reconstruction and fast localization was elaborated. Exploiting the quasi-statistic channel, the 2D-MUSIC algorithm was used to detect the AoAs and AoDs of the paths between the transmitter and the receiver with analog MIMO front-end. The path length can then be estimated via multi-carrier ranging. Based on the AoAs, AoDs, and path lengths estimated by the receiver at different locations, the indoor layout can be reconstructed. With the layout knowledge, we continue to show that the receiver can be localized via the its observed AoAs. The experiment results of this paper demonstrated the feasibility to track the environment and trajectory of mobile devices via mmWave communication signals. With both environment and trajectory information, the communication efficiency may be improved, which is a promising topic for future study.

\bibliographystyle{IEEEtran}
\bibliography{mmWave_sensing_v1_0}
\end{document}